\documentclass[floatfix,twoside, twocolumn,aps]{revtex4}

\usepackage{amssymb,amsmath}
\usepackage{epsfig}

\begin{document}

\title{Analysis of aggregated tick returns: evidence for anomalous diffusion}

\author{Philipp Weber}

\affiliation{
   Institut f\"ur Theoretische Physik, Universit\"at zu
  K\"oln, 50937 K\"oln, Germany\footnote{Electronic address: pw@thp.uni-koeln.de}
  }

\begin{abstract}

In order to investigate the origin of large price fluctuations, we
analyze stock price changes of ten frequently traded NASDAQ stocks in
the year 2002. Though the influence of the trading frequency on the
aggregate return in a certain time interval is important, it cannot
alone explain the heavy tailed distribution of stock price
changes. For this reason, we analyze intervals with a fixed number of
trades in order to eliminate the influence of the trading frequency
and investigate the relevance of other factors for the aggregate
return. We show that in tick time the price follows a discrete
diffusion process with a variable step width while the difference
between the number of steps in positive and negative direction in an
interval is Gaussian distributed. The step width is given by the
return due to a single trade and is long-term correlated in tick
time. Hence, its mean value can well characterize an interval of many
trades and turns out to be an important determinant for large
aggregate returns. We also present a statistical model reproducing the
cumulative distribution of aggregate returns. For an accurate
agreement with the empirical distribution, we also take into account
asymmetries of the step widths in different directions together with
crosscorrelations between these asymmetries and the mean step width as
well as the signs of the steps.

\end{abstract}

\maketitle

The mechanics of stock price changes were studied already more than a
hundred years ago, when Bachelier modelled price movements as a
diffusion process with Gaussian fluctuations \cite{Bachelier00}.
However, empirical studies show that the distribution of returns has
heavy tails \cite{Mandelbrot63, Mantegna95, Lux96, Gopikrishnan98,
Plerou99, Pagan96, Muller98, Fama63, Fama65, Mantegna96, Mantegna97a,
Longin96}, meaning that events with large price changes are much more
probable than in a Gaussian distribution.  In addition, the functional
form of the distribution stays similar if the return is aggregated on
very different time scales from seconds to months, approximating a
Gaussian distribution only if the time scale becomes very
large~\cite{Mandelbrot63, Mantegna95}.

These findings would suggest that stock returns have a L\'evy stable
distribution \cite{Pareto1897, Fama63, Mandelbrot63, Levy37}.  In a
L\'evy flight, the second moment would be divergent and extreme
returns aggregated over a long time would be determined by very large
price jumps on smaller time scales. However, empirical studies find
evidence that the tail of the return distribution follows a power law
with exponent around $\alpha -1 = 3$ so that it does not agree with
the stable Paretian hypothesis \cite{Mantegna95, Lux96,
Gopikrishnan98, Plerou99, Gopikrishnan99, Muller98, Officer72,
Praetz72, Blattberg74, Loretan94, Clark73}.

The cause of the fat tails is currently a subject of great interest
\cite{Gabaix03, WeRo05, Weber06, Farmer03, Plerou00}. Farmer et al.~find
that the distribution of returns due to a single trade (tick returns)
is similar to the distribution of returns aggregated on longer time
scales with the same tail exponent~\cite{Farmer03}.  Although the tail
exponent is outside the L\'evy regime $0 < \alpha - 1 < 2$, they argue
that similar to a L\'evy flight both distributions are caused by the
same microscopic mechanism, so that large aggregate returns are due to
single exceptionally large tick returns.  Plerou et al.~describe the
price movements as a diffusion process with a fluctuating diffusion
constant and relate the distribution of aggregate returns to the
distribution of the variance of the tick returns~\cite{Plerou00}.

In the present paper, we investigate the transition from tick returns
to returns aggregated in intervals with a larger number of trades. It
is well documented (e.g.~in \cite{Jo+94, Ea+97}) that the
number of trades in a time interval is an important determinant of the
aggregate return. However, the trading frequency alone cannot account
for the observed fat tailed distribution of aggregate returns
\cite{Plerou00}. Thus, we remove the direct influence of the trading
frequency by analyzing intervals with a constant number of trades so
that effects due to other quantities like the tick return size are
more clearly visible.

We study how each aggregate return is actually built from the basic
quantities involved in the process, and thus examine the mechanism
leading to large price fluctuations. According to the central limit
theorem, independent tick returns would in aggregation lead to
Gaussian distributed returns. However, we find that the tick return
size is long-term correlated in tick time (compare~\cite{Ding93,
Granger95, Ding96, Andersen97, Liu97, Cizeau97, Cont98, Pasquini99,
Liu99}), so that the conditions of the central limit theorem are not
fulfilled. Thus, the mean tick return size can well characterize an
interval of many trades and its fluctuations lead to the non-Gaussian
behavior of the aggregate return. In this picture, large aggregate
returns do not occur because of a few very large tick returns, but
rather when the average tick return is large, so that even Gaussian
fluctuations in the direction of the trades can lead to aggregate
return larger than in a Gaussian distribution.

The remainder of this paper is organized as follows: section I shows
our model for the price diffusion process, in section II we describe
the data set used for this study, section III shows the influence of
the tick return size on the aggregate return while section IV focusses
on the influence of differences in the direction of tick returns
(number difference). Section V compares the number difference and the
flow of market orders and in section VI we present a statistical model
which approximates the distribution of aggregate returns. We conclude
with a discussion of our results in section VII.

\section{Model}

We study intervals with a fixed number of $N=100$ trades. If the price
of a stock before the $i$-th trade is $s_i$, we define the return due
to a single trade, the tick return, as
\begin{equation}
\delta g_i = ln(s_{i+1}) - ln(s_{i}) \ \ .
\label{tick_returns.eq}
\end{equation}

The interval $I_j$ contains all $N$ trades with index $i$ between $jN$
and $(j+1)N$, so the aggregate return $G_j$ is given by the sum over
all $\delta g_i$ with $i \epsilon I_j$:
\begin{equation}
G_j = \sum_{i \epsilon I_j}\delta g_i \ \ .
\label{aggregate_returns.eq}
\end{equation}
We want to discuss two special cases in order to analyze the mechanism
leading to large aggregate returns $G_j$.  In the first case, $G_j$ is
dominated by one (or a few) extremely large $\delta g_{i_0}^{max}$, so
that
\begin{equation}
G_j = \delta g_{i_0}^{max} + \sum_{i \epsilon I_j, i \ne i_0}\delta g_i \approx \delta g_{i_0}^{max} \ \ .
\label{g_i_max.eq}
\end{equation}
Thus, $G_j$ becomes large if $\delta g_{i_0}^{max}$ is exceptionally
large.

In the second case, we assume that there is no extremely large tick
return dominating the aggregate return, so that we focus on the
average size $\Delta g_j$ of the non-zero tick returns, which is
defined by
\begin{equation}
\Delta g_j = \frac{1}{n_j}\sum_{\delta g_i \neq 0, i \epsilon I_j}|\delta g_i|
\label{delta_g.eq}
\end{equation}
Here, $n_j$ is the number of $\delta g_i \ne 0$ in the interval $I_j$.
Neglecting assymetries in the $\delta g_i$, we can replace all $\delta
g_i \ne 0$ by $\rm{sign}(\delta g_i) \Delta g_j$ and approximate the
aggregate return by
\begin{equation}
G_j \approx \Delta g_j \sum_{\delta g_i \neq 0, i \epsilon I_j} \rm{sign}(\delta g_i) = \Delta g_j \Delta N_j
\label{g_i.eq}
\end{equation}
where $\Delta N_j = \sum_{\delta g_i \neq 0, i \epsilon I_j}
\rm{sign}(\delta g_i)$ is called number difference. Similarly, $G_j$ can
be described as a diffusion process with
\begin{equation}
\left < G_j^2 \right > \approx D_j N
\label{diffusion.eq}
\end{equation}
where the diffusion constant $D_j = \frac{n_j}{N}\Delta g_j^2$ varies
due to the varying step width $\Delta g_j$ and the number $n_j$ of
non-zero tick returns.

In the approximation given by Eq.~\ref{g_i.eq}, we can study the
influence of the mean size of the tick returns as well as asymmetries
in their direction. A large aggregate return can occur if the price
moves more often in one direction than in the other. Thus, with large
temporary correlations between the signs, even small tick returns
could compose a large $G_j$. On the other hand, if $\Delta g_j$ is
larger, even a small asymmetry in the signs can lead to a large
return.

The two approximations given in Eq.~\ref{g_i_max.eq} and
Eq.~\ref{g_i.eq} are analyzed in sections III and IV of this paper,
but in section VI we also consider the error term neglected in
Eq.~\ref{g_i.eq}.  An exact formulation writes
\begin{equation}
G_j = \Delta g_j \Delta N_j + \frac{2 n_j^+ n_j^-}{n_j}(\Delta g_j^+ - \Delta g_j^-)
\label{g_j_exact.eq}
\end{equation}
where $\Delta g_j^+$ and $\Delta g_j^-$ are the average tick returns
in positive and negative direction while $n_j^+$ and $n_j^-$ are the
numbers of non-negative tick returns in positive and negative
direction.

\section{Data analysis}

We analyzed order book data of the year 2002 from Island ECN for the
ten most frequently traded stocks~\cite{ticker}. Since the Island ECN
is a secondary market where only part of the whole stock volume is
traded, we also studied the index fund QQQ which was mainly traded via
Island until September 2002. Since our results for the ten stocks and
QQQ are similar, we find no evidence that secondary market
characteristics of Island affect our analysis negatively.

In an electronic market place like Island, people can place limit
orders to buy or to sell at a given or better price (limit price),
which is specified in the order. These orders are stored in the order
book and they are only executed when the actual stock price reaches
the limit price.  A trade is initiated by a market order indicating that
someone wants to buy or sell immediately at the best available
price. Such a market order executes the limit orders offering the best
prices until the number of shares specified in the market order is
traded.

Our dataset contains information about every limit order so that we
are able to reproduce the market situation at each instant of time. We
combine those limit order executions with identical time stamps as
they reflect the same market order.  Therefore, we can analyze the
impact of each single market order on the price. In this analysis, the
price is defined as the mid-quote price which is the mean of the best
available buy (bid) and sell (ask) limit prices (quotes).  We study
intervals with a fixed number of $N=100$ market orders and have
approximately 100,000 intervals in our dataset for ten stocks. Thus,
on average a 100 trade interval corresponds to about 10 minutes, but
the trading frequency fluctuates strongly so that 100 trades can
correspond to time intervals with very different lengths.

We determine the mid-quote price $s_i$ just before the execution of
the $i$-th market order. Since most trades change the price just by
the size of the gap between the best and the second best limit price
\cite{Farmer03}, the tick return $\delta g_i$ corresponds to the gap
size. We note that the price can (and often does) change between two
consecutive market orders due to placement or cancellation of limit
orders so that $\delta g_i$ does not provide a direct estimate of the
gap size. We normalize the tick returns $\delta g_i$ by the standard
deviation of the aggregate return $G_j$ for each stock individually so
that we can combine the results for different stocks.

%
\begin{figure}
  \centerline{ \epsfig{file=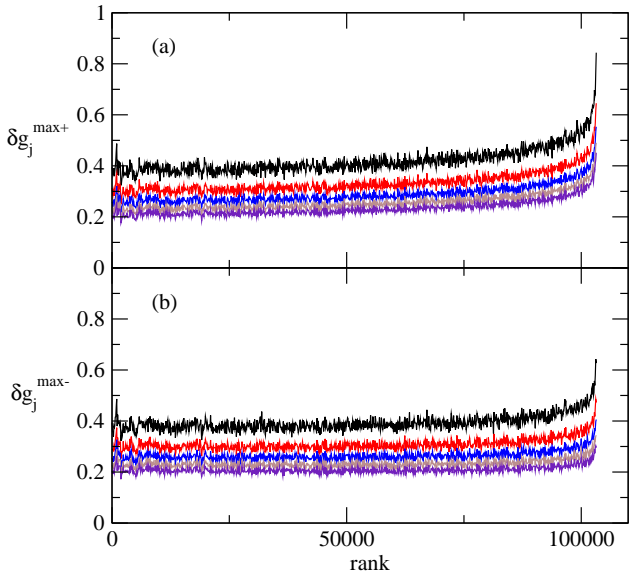,width=8.3cm}}

  \caption{(Color online) Five largest price changes (a) $\delta
  g_{j}^{max+}$ and (b) $\delta g_{j}^{max-}$ due to a single trade
  with (a) the same and (b) the opposite sign as the aggregate return
  in that 100 tick interval, plotted against the rank of the
  corresponding aggregate return $|G_j|$ for the combined data of ten
  Nasdaq stocks in 2002 (smoothed by averaging over 100
  intervals). For large $|G_j|$, the size of the $\delta g_{j}^{max+}$
  increases by a factor of two while the increase in the $\delta
  g_{j}^{max-}$ is slightly smaller. The sum over all five $\delta
  g_{j}^{max+}$ reaches more than 3 standard deviations for intervals
  with extremely large $|G_j|$, but the fluctuations in the opposite
  direction are almost equally large. }

  \label{g_plus_max_12345_rank_other_avg.fig}
\end{figure}
%

\section{Influence of the size of tick returns}

First, we investigate the question whether large tick returns caused
by large gaps in the order book can be responsible for large aggregate
returns. To this end, we start with the approximation shown in
Eq.~\ref{g_i_max.eq} where a few extremely large tick returns
(corresponding to some very large gaps in the order book) lead to a
very large aggregate return $G_j$. In order to test this hypothesis,
we analyzed the five largest tick returns $\delta g_{j}^{max+}$ with
the same sign as the aggregate return $G_j$ (i.e. the five largest
positive $\delta g_i$ if $G_j>0$ and the five largest negative $\delta
g_i$ for $G_j<0$) in each time interval. To this end, we sort the
intervals by $|G_j|$ and plot the $\delta g_{j}^{max+}$ against the
rank of the interval according to its return $|G_j|$.

Fig.~\ref{g_plus_max_12345_rank_other_avg.fig}(a) shows the values of
these $\delta g_{j}^{max+}$ in intervals with small $G_j \approx 0$ on
the left while the values for large returns exceeding five standard
deviations can be found on the right. Since there are large
fluctuations in the data, we smoothed the curves by averaging over 100
intervals. The $\delta g_{j}^{max+}$ grow by a factor of two between
small and very large returns $|G_j|$. When aggregated, these five
largest $\delta g_{j}^{max+}$ can reach about three standard
deviations, which is almost half of the largest aggregate returns.

%
\begin{figure}
  \centerline{ \epsfig{file=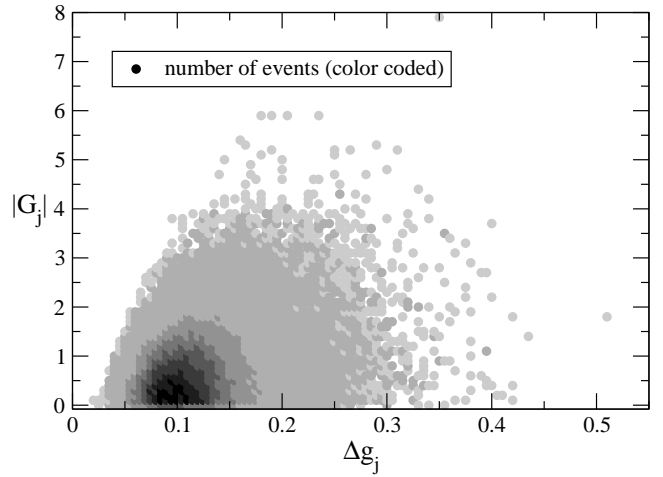,width=8.5cm}}
  \caption{Density plot of the 100-trade-return $|G_j|$ of ten Nasdaq
  stocks against the average return of a single trade $\Delta g_j$ for
  each interval. The Points are coded from light grey to black
  indicating the number of events from 1 to more than 500. A linear
  regression has only a small correlation coefficient $R^2=0.07$.}
  \label{regression_g_dg.fig}
\end{figure}
%

In Fig.~\ref{g_plus_max_12345_rank_other_avg.fig}(b), we plot the five
largest tick returns $\delta g_{j}^{max-}$ with the opposite direction
as the aggregate return against their rank. The $\delta g_{j}^{max-}$
behave similarly to the $\delta g_{j}^{max+}$, though the increase for
large aggregate returns is slightly weaker. However, even for the
largest aggregate returns the difference between the $\delta
g_{j}^{max+}$ and $\delta g_{j}^{max-}$ is rather small, so that there
are also large tick returns reducing the aggregate return.  

Our results suggest that large aggregate returns are not the result of
single exceptionally large tick returns since very large tick returns
occur in both directions and cancel each other out. In the following,
we want to focus not on the extreme tick returns, but on the influence
of their mean value. More precisely, we analyze Eq.~\ref{g_i.eq} and
the mean tick return $\Delta g_j$ of all non-zero $|\delta g_i|$ in
the interval $I_j$ as defined in Eq.~\ref{delta_g.eq}. A density plot
of $|G_j|$ against $\Delta g_j$ is shown in
Fig.~\ref{regression_g_dg.fig}. It seems that extremely large returns
$G_j$ correspond to larger average tick returns $\Delta g_j$, but the
broad distribution suggests that the explanatory power of $\Delta g_j$
alone for the aggregate return $G_j$ is small, which is confirmed by
the low correlation coefficient $R^2=0.07$ of a linear regression.

%
\begin{figure}
  \centerline{ \epsfig{file=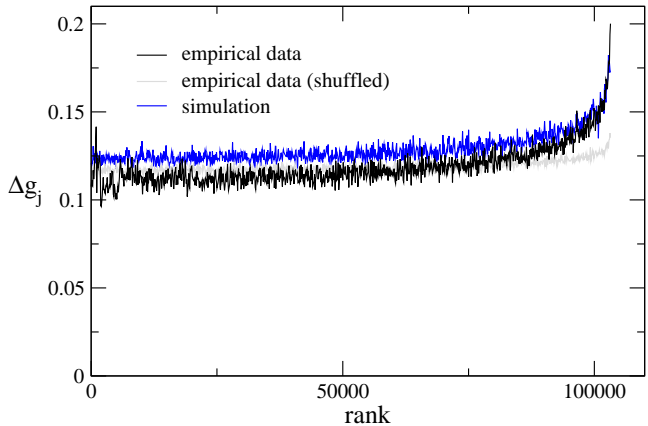,width=8.5cm}}
  \caption{(Color online) Black curve: average tick return $\Delta
  g_j$ of ten Nasdaq stocks plotted against the rank of the
  corresponding aggregate return $|G_j|$, smoothed by averaging over
  100 intervals. Going from the smallest returns $|G_j| \approx 0$ to
  returns larger than 5 standard deviations, $\Delta g_j$ increases by
  a factor of two. Light grey curve: after shuffling the tick returns
  for each stock, the same curve is only slightly increased for the
  largest aggregate returns, the effect is much smaller than for the
  original data. Blue curve (or dark grey): the simulation according
  to the statistical model discussed in section VI shows a similar
  behaviour as the empirical data, but in the simulation $\Delta g_j$
  is a little larger than the empirical one except for the largest
  $|G_j|$ where the simulated $\Delta g_j$ is slightly smaller than
  the empirical one.}

  \label{dg_rank.fig}
\end{figure}
%

In order to clarify the relation between the extreme values of $|G_j|$
and $\Delta g_j$, we sort the intervals by $|G_j|$ and plot $\Delta
g_j$ against the rank of the interval according to its return
$|G_j|$. In Fig.~\ref{dg_rank.fig} (black curve), we see that large
returns $|G_j|$ coincide with larger tick returns as $\Delta g_j$
changes by a factor of two from very low aggregate returns to large
returns of several standard deviations. In comparison with the largest
tick returns $\delta g_{j}^{max+}$ shown in
Fig.~\ref{g_plus_max_12345_rank_other_avg.fig}, the change of a factor
of two is similar, but the mean $\Delta g_j$ is two to four times
smaller than the largest tick returns.

\begin{figure}
  \centerline{ \epsfig{file=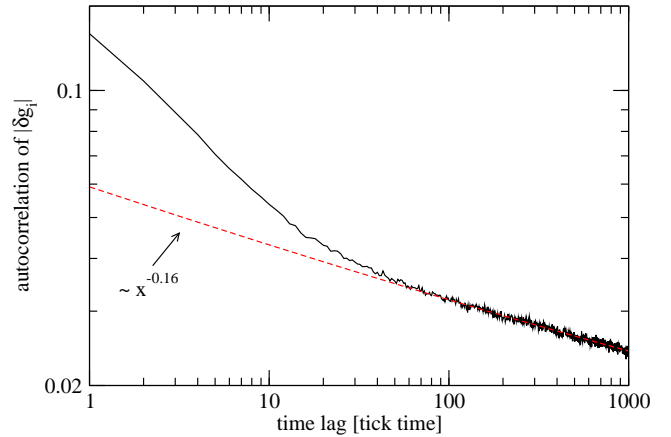,width=8.5cm}}
  \caption{Autocorrelation function of the absolute value of the tick
  return $|\delta g_i|$ averaged over the data of ten Nasdaq stocks in
  2002. The function shows a power law decay in tick time proportional
  to $\Delta i^{-0.16}$ for large $\Delta i$.}

  \label{autocorr.fig}
\end{figure}

This finding can be explained by the presence of autocorrelations in
the time series of $\delta g_i$, which can be illustrated when we
shuffle the data for each stock by exchanging each tick return with
another randomly chosen tick return. The light grey curve in
Fig.~\ref{dg_rank.fig} shows that for shuffled data $\Delta g_j$
increases only marginally for large aggregate returns, suggesting that
autocorrelations of the tick returns have a strong influence on the
mean tick return size $\Delta g_j$. Indeed, we find that the absolute
values $|\delta g_i|$ of the tick return are long-range correlated in
tick time with a correlation function decaying like $\Delta i^{-0.16}$
for large time lags $\Delta i = |i_1 - i_2|$, as shown in
Fig.~\ref{autocorr.fig}. If these correlations are destroyed by
shuffling, in each interval of 100 trades only a few large tick
returns remain so that the average over these 100 tick returns
approximates the global mean of all tick returns in the data set.

In contrast, in the empirical unshuffled data correlations lead to
intervals where many tick returns are large, so that the average tick
return size is also large. The average tick return size $\Delta g_j$
can well characterize the interval only because these autocorrelations
exist. It turns out that the increase of $\Delta g_j$ by a factor of
two is the main effect where the original empirical data deviates
significantly from shuffled data. Hence, we suggest that fluctuations
of the tick return size are responsible for the non-Gaussian
fluctuations of the aggregate return.

%
%

Using Eq.~\ref{g_i.eq}, we can estimate whether the change by a factor
of two of the average tick return alone is enough to explain large
aggregate returns $G_j$ of more than five standard deviations. To this
end, we focus on the intervals with the 50 largest aggregate returns
ranging from approximately 4 to almost 8 standard deviations. Here, we
find that $\Delta g_j$ fluctuates between 0.14 and 0.35. Assuming
uncorrelated returns, $\Delta N_j$ should be of the order $\sqrt{N}
\approx 10$ if each trade would lead to a price change, but normal
fluctuations could well lead to $\Delta N_j$ twice as large as
$\sqrt{N}$, so that large tick returns together with fluctuations in
the number difference could explain the large aggregate returns we
find in our data set.

Thus, we find that in intervals with 100 trades large $|G_j|$ do not
mainly depend on single extremely large tick returns. It rather turns
out that correlations between the tick returns lead to large average
tick returns $\Delta g_j$ in an interval, and the fluctuations of
$\Delta g_j$ can account for the non-Gaussian distribution of the
aggregate returns.

\section{Number difference}

%
\begin{figure}
  \centerline{ \epsfig{file=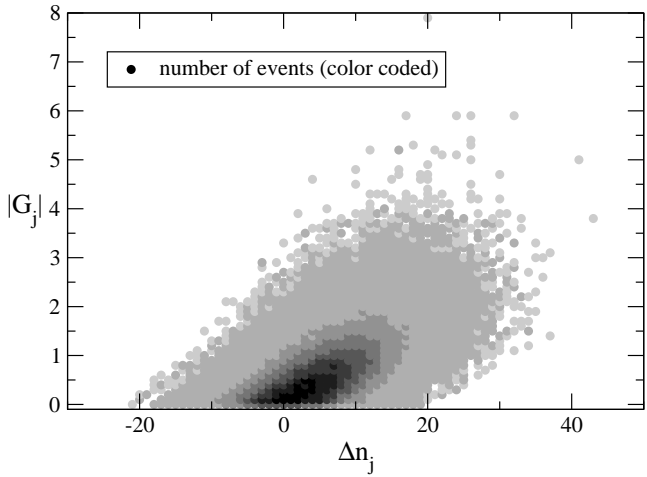,width=8.5cm}}
  \caption{Density plot of the aggregate return $|G_j|$ against the
  difference $\Delta n_j$ between the number of tick returns with the
  same and with the opposite direction as the aggregate return, for
  ten Nasdaq stocks. The points are coded from light grey to black
  indicating the number of events from 1 to more than 600. A linear
  regression has a large correlation coefficient $R^2=0.32$.}

 \label{regression_g_dn.fig}
\end{figure}
%

The diffusion process of aggregate returns is not only influenced by
the step width (i.e.~the tick return size), but also by the direction
of the steps. Therefore, we now analyze the influence of the number
difference $\Delta N_j$ in Eq.~\ref{g_i.eq}. In order to treat
positive and negative aggregate returns in the same analysis, it is
useful to replace $\Delta N_j$ by the sign-adapted number difference
\begin{equation}
\Delta n_j = \rm{sign}(G_j)\Delta N_j \ \ .
\label{dn.eq}
\end{equation}
A positive value of $\Delta n_j$ indicates that the price tends to
move in one specific direction leading to an aggregate return with the
same sign. $\Delta n_j$ can be negative if there are a few large tick
returns determining the direction of the aggregate return, but also
many small tick returns with the opposite direction which do not
affect the aggregate return very much. Fig.~\ref{regression_g_dn.fig}
shows a density plot of the aggregate return $|G_j|$ against the
sign-adapted number difference $\Delta n_j$. A linear regression
yields an $R^2$ of $0.32$, a large correlation coefficient confirming
the visual impression that $\Delta n_j$ and $|G_j|$ are strongly
connected. We can also see that $\Delta n_j$ is mostly positive for
large returns $G_j$, so that each large price change is accompanied by
a certain sign-adapted number difference $\Delta n_j$.

\begin{figure}
  \centerline{ \epsfig{file=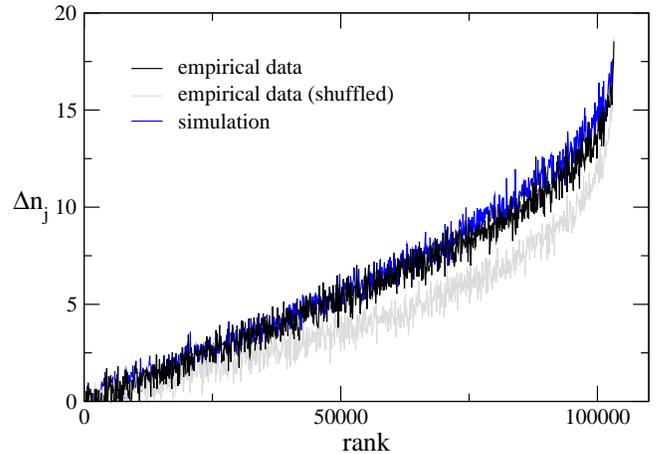,width=8.5cm}}
  \caption{(Color online) Black curve: the sign-adapted number
  difference $\Delta n_j$ is plotted against the rank according to the
  aggregate return $|G_j|$ for 10 Nasdaq stocks, smoothed by averaging
  over 100 intervals. $\Delta n_j$ grows from zero to 18. The relation
  between $\Delta n_j$ and the rank seems to be linear except for the
  largest 15\% of the aggregate returns. A simulation (blue curve (or
  dark grey)) using a normal distribution for $\Delta N_j$ leads to
  nearly the same dependance on the rank. For shuffled data (light
  grey curve), the curve is slightly flatter, but the difference is
  not large.}
  
\label{dn_rank.fig}
\end{figure}

We now plot in Fig.~\ref{dn_rank.fig} $\Delta n_j$ against the rank
according to $|G_j|$. We find that except for the largest
(approximately 15\%) of the aggregate returns, $\Delta n_j$ grows
linearly with the rank while in Fig.~\ref{dg_rank.fig} $\Delta g_j$
remained almost constant in that region. For the largest ranks,
$\Delta n$ increases more rapidly, so that all in all the smoothed
curve (averaged over 100 intervals) grows from 0 to 18 between very
small and extremely large aggregate returns. Thus, in intervals with
very large returns there are approximately 18 trades pushing the price
in one direction (assuming that all other trades cancel each other
out), so that even with rather small tick returns this can lead to
large returns in aggregation. Focusing on the 50 largest $G_j$, we
find that $\Delta n_j$ ranges from 4 to 41, clearly above the expected
standard deviation of 10 when assuming uncorrelated returns and $n_j =
N$.

%
%

Thus, the fluctuations of $\Delta n_j$ around the mean value are
crucial for getting large aggregate returns. The number difference
seems to be the main mechanism affecting the aggregate return since it
changes much more drastically than the tick return size when the
aggregate return increases. On the other hand, when we compare the
results to the analysis with shuffled data (light grey curve in
Fig.~\ref{dn_rank.fig}), it turns out that this effect is very similar
to what happens with random price changes. Hence, the basic movement
of the aggregate return seems to depend mostly on the number
difference, but the non-Gaussian large aggregate price changes only
occur if the tick returns are large.

\section{Market order signs and direction of tick returns}

It is known that the signs of market orders are strongly correlated
\cite{Bou+04, LiFa04} which means that there is a large probability that
a buy market order will be followed by another buy market order. Thus,
it is probable that large number differences in the direction of tick
returns are caused by large numbers of equally signed market
orders. In order to analyze the relation between the number difference
and the market order flow, we define the difference $\Delta n^{m}_j$
between the number $n^{m+}_j$ of market orders with the same direction
as $G_j$ and the market orders with opposite direction $n^{m-}_j$:

\begin{equation}
\Delta n^{m}_j = n^{m+}_j - n^{m-}_j \ \ .
\label{dnm_eq}
\end{equation}

%
\begin{figure}
  \centerline{ \epsfig{file=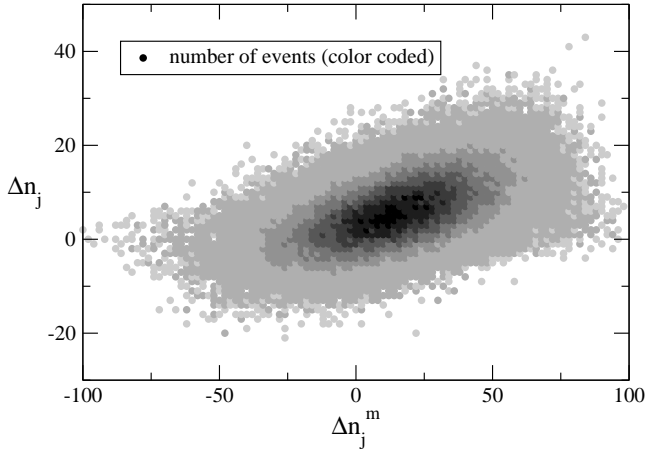,width=8.5cm}}

  \caption{Comparison between sign-adapted number difference $\Delta
  n_j$ and market order difference $\Delta n_{j}^{m}$ for ten Nasdaq
  stocks.  The Points are coded from light grey to black indicating
  the number of events from 1 to more than 200. The correlation
  coefficient of a linear regression yields $R^2=0.29$, thus there is
  a strong connection between the two quantities. On the other hand,
  the events scatter widely so that small $\Delta n$ are often linked
  with large $\Delta n_{j}^{m}$ and vice versa.}
  \label{scatter_dn_dnm.fig}
\end{figure}
%

In Fig.~\ref{scatter_dn_dnm.fig} we plot the sign-adapted number
difference $\Delta n_j$ against the market order difference $\Delta
n^{m}_j$. We find a strong correlation between $\Delta n_j$ and
$\Delta n^{m}_j$, a linear regression yields a correlation coefficient
$R^2$ of 0.29. However, there are also large fluctuations suggesting
that the number difference is also due to order book dynamics, namely
limit order placement and cancellation as well as asymmetries in the
order book. A model for price formation due to these quantities was
recently proposed by Mike and Farmer
\cite{MiFa05}.

\section{Distribution of aggregate returns and a statistical model}

In the first part of this paper, we analyzed the mechanism leading to
large aggregate returns and showed that the varying step width $\Delta
g_j$ accounts for the non-Gaussian behavior of the diffusion process
of price movements. Now we want to use our results in a statistical
model and reproduce the cumulative distribution function of the
absolute value of the aggregate return $|G_j|$.

The model given by Eq.~\ref{g_i.eq} belongs to the well-known class of
stochastic volatility models (see e.g.~\cite{Ca+97}) consisting of a
noise term multiplied by a time-dependent volatility giving the
magnitude of the fluctuations.  In the present paper, the model is
based on a microscopic description of the price process, so that we
can fit the microscopic quantities determining the aggregate return in
order to estimate the parameters of the model. In this approach the
model is parameter-free in the sense that there are no parameters
fitting the aggregate returns directly, though we fit the
distributions of its determinants like the step width $\Delta g_j$ and
the number difference $\Delta N_j$. We also discuss corrections to the
model by including the tick return asymmetries according to
Eq.~\ref{g_j_exact.eq}.

%
\begin{figure}
  \centerline{ \epsfig{file=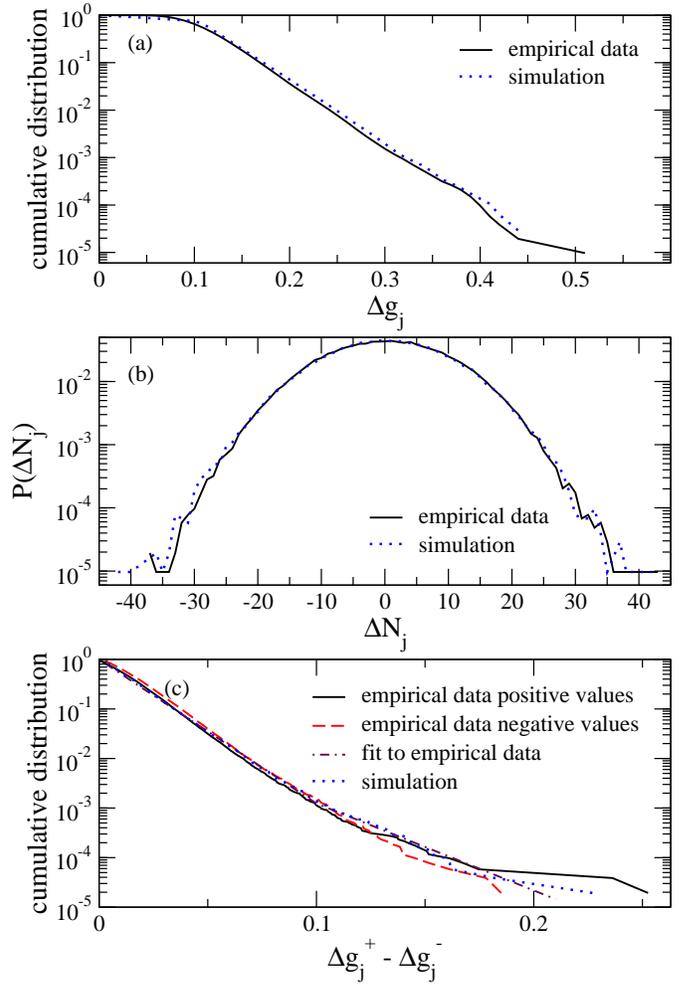,width=8.8cm}}

  \caption{(Color online) Estimation of the parameters for the
  simulation (results shown as dotted lines) from empirical data for
  ten Nasdaq stocks. (a) The tail of the cumulative distribution of
  $\Delta g_j$ (line) can be well fitted with $P(x>\Delta g_j) = e^{-a
  (x-x_0)/\Delta\bar{g}}$ where $\Delta\bar{g}\approx 0.12$ is the
  average of all $\Delta g_j$ and the parameters are $a=3.6$ and
  $x_0=0.094$. For $\Delta g_j \lesssim x_0$ the limited tick size
  leads to a plateau. (b) The probability distribution of $\Delta N_j$
  (line) follows in good approximation a normal distribution with mean
  0.24 and standard deviation 9.0. (c) As a rough approximation, the
  average of the cumulative distribution of the positive (line) and
  negative (dashed line) values of $\Delta g_j^+ - \Delta g_j^-$ are
  parameterized proportional to two exponential functions
  $e^{-a_{1,2}x/\Delta\bar{g}}$ for $|\Delta g_j^+ - \Delta g_j^-|
  \lessgtr 0.1$, with $a_1=8.0$ and $a_2=4.8$ (dashdotted line). The
  simulation (dotted line) uses the adapted $a_1=9.0$ and $a_2=2.0$ in
  order to compensate the change in the distribution after taking into
  account $\left<\Delta g_j^+ - \Delta g_j^- \right>_{\Delta g_j
  \Delta N_j}$ .}

 \label{distribution_dg_dn_dgpm.fig}
\end{figure}
%

We first analyze the distributions of $\Delta g_j$ and $\Delta
N_j$. Fig.~\ref{distribution_dg_dn_dgpm.fig}(a) shows the cumulative
distribution of $\Delta g_j$ in a log-linear plot. The approximately
straight line suggests that the tail follows an exponential
distribution which can be well fitted with $P(x>\Delta g_j) = e^{-a
(x-x_0)/\Delta\bar{g}}$ where $\Delta\bar{g}\approx 0.12$ is the
average of all $\Delta g_j$ and the parameters are $a=3.6$ and
$x_0=0.094$. In the region of the smallest values of $\Delta g_j
\lesssim x_0$ the limited tick sizes of the different stocks lead to a
plateau. In section IV we already found evidence that $\Delta N_j$
behaves similarly to uncorrelated data since in Fig.~\ref{dn_rank.fig}
the shuffled data shows almost the same dependence on the rank of the
corresponding $|G_j|$.  Figure
\ref{distribution_dg_dn_dgpm.fig}(b) shows that indeed $\Delta N_j$
can be well described by a Gaussian noise with mean 0.24 and standard
deviation 9.0.

%
\begin{figure}
  \centerline{ \epsfig{file=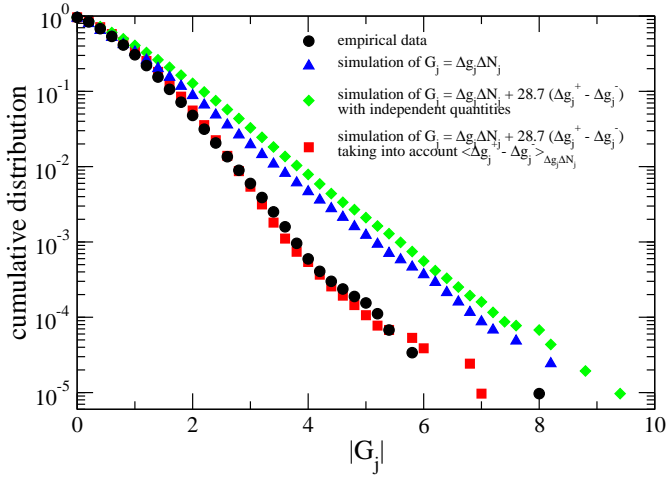,width=8.8cm}}

  \caption{(Color online) Cumulative distribution of the empirical
  aggregate return (circles) obtained from ten Nasdaq stocks in
  comparison with different simulations. A simulation of
  Eq.~\ref{g_i.eq} (triangles) leads to a reasonable approximation of
  the empirical data, but it overestimates the probability of large
  returns. It becomes a little broader if we add the tick return
  asymmetry $\Delta g_j^+ - \Delta g_j^-$ according to
  Eq.~\ref{g_j_exact.eq} and simulate independent quantities
  (diamonds).  The simulation (squares) matches the empirical data
  very well if we incorporate correlations by generating $\Delta g_j^+
  - \Delta g_j^-$ according to the conditional expectation value
  $\left<\Delta g_j^+ - \Delta g_j^- \right>_{\Delta g_j \Delta N_j}$
  .}

  \label{distribution_g.fig}
\end{figure}
%

In order to analyze the accuracy of the approximation given in
Eq.~\ref{g_i.eq}, we simulate two independent time series according to
the fitted functions for $\Delta g_j$ and $\Delta N_j$ and build the
aggregate return $G_j$ as the product of $\Delta g_j$ and $\Delta
N_j$. In Figure \ref{distribution_g.fig} we can compare the
empirically found cumulative distribution of aggregate returns $|G_j|$
(circles) to the results of this simulation (triangles). The
simulation of Eq.~\ref{g_i.eq} leads to a reasonable agreement with
the actual aggregate return, but it overestimates the probability of
large aggregate returns. We note that the parameters of the simulation
are completely determined by the empirically found distributions of
$\Delta g_j$ and $\Delta N_j$, so that in this sense the simulation of
$|G_j|$ has no free parameters.

In the following, we want to address the remaining deviations of the
simulation from the empirical data. Eq.~\ref{g_j_exact.eq} gives an
exact formula for $G_j$ and provides a good parameterization for the
error term which reads
\begin{equation}
G_j - \Delta g_j \Delta N_j = \frac{2 n_j^+ n_j^-}{n_j}(\Delta g_j^+ - \Delta g_j^-)
\label{error.eq}
\end{equation}
We find that the term $2 n_j^+ n_j^-/n_j$ has no systematic influence
on the aggregate return since it shows almost no dependence on the
rank according to the aggregate return. In the following, we thus
approximate it by its average value $\left <2 n_j^+ n_j^-/n_j\right >
= 28.7$, so that the error term is determined by the asymmetries
$\Delta g_j^+ - \Delta g_j^-$ in the mean tick return size.

The cumulative distribution of $\Delta g_j^+ - \Delta g_j^-$ is shown
in Fig.~\ref{distribution_dg_dn_dgpm.fig}(c). The main part of the
distribution could be well fitted by an exponential function, but in
the tail the distribution becomes broader. Thus, we add the term with
$\Delta g_j^+ - \Delta g_j^-$ to our simulation by creating a third
independent time series according to the empirical distribution of
$\Delta g_j^+ - \Delta g_j^-$.  Fig.~\ref{distribution_g.fig}
(diamonds) shows that this leads to an even broader distribution of
the aggregate return. Since the difference to the distribution
according to Eq.~\ref{g_i.eq} is small, the tick return asymmetry seems
to have only a small influence on the aggregate return.

A more accurate agreement with the empirical data can be obtained by
taking into account correlations between the quantities involved in
the process. The correlation coefficients between them are shown in
the following table where the correlations between the absolute values
are shown in brackets:
%
\begin{center}
\begin{table}[hbt]
\begin{tabular}{|c|c|c|c|}
  \hline
 &   &   &  \\[-0.35cm]
 & $\Delta N_j$ & $\Delta g_j^+ - \Delta g_j^-$ & $\Delta g_j \Delta N_j$ \\
  \hline
 &   &   &  \\[-0.35cm]
$\Delta g_j$ & -0.02 (-0.07) & -0.01 (0.37) & -0.01 (0.34)   \\[0.1cm]
$\Delta N_j$ & 1 & -0.35 (0.01) & 0.95 (0.87)  \\[0.1cm]
$\Delta g_j \Delta N_j$ & 0.95 (0.87) & -0.41 (0.02) & 1   \\[0.1cm]
  \hline
\end{tabular}
\label{size.tab}
\end{table}
\end{center}
%
$\Delta g_j$ and $|\Delta N_j|$ show slightly negative correlations
which might suggest that people act more cautiously when large tick
returns indicate a low liquidity. In these times, traders try not to
place too many consecutive orders with the same sign because they know
that it could lead to a large price change and increased trading
costs. Furthermore, the strong anti-correlations between $\Delta N_j$
and $\Delta g_j^+ - \Delta g_j^-$ also indicate cautious traders: If
there are large asymmetries, so that e.g.~the positive tick returns
are much larger than the negative ones, people tend to use the higher
liquidity in negative direction so that in these times they sell more
often than they buy. For an analysis of the relation between liquidity
imbalance and market efficiency, see e.g.~\cite{Fa+06}. The large
correlations between $\Delta g_j$ and $|\Delta g_j^+ - \Delta g_j^-|$
show that we can expect large variations of the tick return in
positive and negative direction when the tick return is in general
large.

%
\begin{figure}
  \centerline{ \epsfig{file=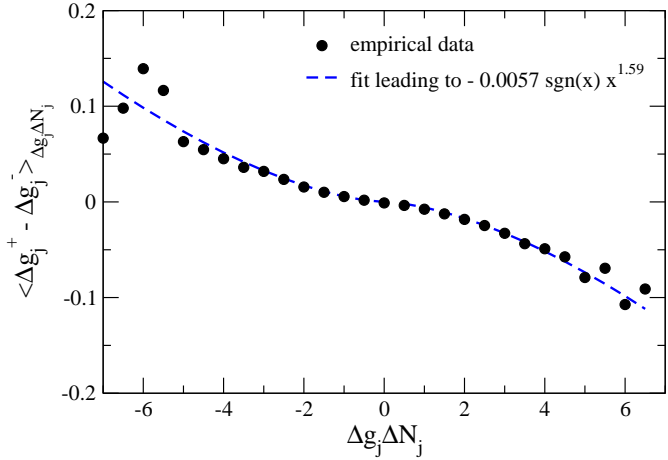,width=8.8cm}}

  \caption{(Color online) Conditional expectation value $\left<\Delta
  g_j^+ - \Delta g_j^- \right>_{\Delta g_j \Delta N_j}$ plotted
  against $\Delta g_j \Delta N_j$ (circles), obtained from the data of
  10 Nasdaq stocks. A fit leads to $\left<\Delta g_j^+ - \Delta g_j^-
  \right>_{\Delta g_j \Delta N_j} \approx - 0.0057 \cdot sgn(\Delta
  g_j \Delta N_j)\cdot (\Delta g_j \Delta N_j)^{1.59}$ (dashed
  line). The tick return asymmetry $\Delta g_j^+ - \Delta g_j^-$ is
  strongly correlated with the mean tick return size $\Delta g_j$ and
  strongly anti-correlated with the number difference $\Delta
  N_j$. Using the conditional expectation value in the simulation
  incorporates these correlations which allows for the reproduction of
  the distribution of aggregate returns.}

  \label{condexp_dgpm_dgdn.fig}
\end{figure}
%

We now want to incorporate correlations in our simulation. The
strongest non-trivial correlations appear between $\Delta g_j \Delta
N_j$ and $\Delta g_j^+ - \Delta g_j^-$ including also some of the
correlations between $\Delta g_j^+ - \Delta g_j^-$ and $\Delta g_j$ as
well as $\Delta N_j$. However, it turns out that the conditional
expectation value $\left<\Delta g_j^+ - \Delta g_j^-\right>_{\Delta
g_j \Delta N_j}$ is non-linear, as seen in
Fig.~\ref{condexp_dgpm_dgdn.fig} (circles) where it is plotted against
$\Delta g_j\Delta N_j$. The function can be well fitted by
$-\rm{sgn}(x) \alpha |x|^{\beta}$ with $\alpha=0.0057$ and
$\beta=1.59$ (dashed line).

In order to incorporate this conditional expectation value into the
simulation, we first create three independent time series for $\Delta
g_j$, $\Delta N_j$, and $\Delta g_j^+ - \Delta g_j^-$. Then, for
each $j$ we add the conditional expectation value $\left<\Delta g_j^+
- \Delta g_j^-\right>_{\Delta g_j \Delta N_j}$ to $\Delta g_j^+ -
\Delta g_j^-$, according to the value of $\Delta g_j \Delta N_j$ for
that $j$. This method leads to a different distribution for $\Delta
g_j^+ - \Delta g_j^-$ than the initial one, so that we can not anymore
generate $\Delta g_j^+ - \Delta g_j^-$ from the unconditional
empirical distribution. As a rough approximation, we parameterize this
distribution by two exponential functions
$e^{-a_{1,2}x/\Delta\bar{g}}$ for $\Delta g_j^+ - \Delta g_j^-
\lessgtr 0.1$. Then, we adapt the factors in the exponent in such a way
that the resulting unconditional distribution fits the empirical one
(a fit to the empirical distribution yields $a_1=8.0$ and $a_2=4.8$,
for the simulation we use the adapted $a_1=9.0$ and $a_2=2.0$, compare
Fig.~\ref{distribution_dg_dn_dgpm.fig}(c)). The resulting distribution of
$G_j$ does not depend very much on the exact values of $a_{1,2}$.

The effect of the correlations represented by the conditional
expectation value $\left<\Delta g_j^+ - \Delta g_j^-\right>_{\Delta
g_j \Delta N_j}$ is very large and leads to a cumulative distribution
of $|G_j|$ (squares in Fig.~\ref{distribution_g.fig}) very similar to
the empirical one (circles). It is worth noting that now the largest
events are not anymore necessarily the ones with the largest values of
$\Delta g_j \Delta N_j$. Due to the anti-correlations expressed in
$\left<\Delta g_j^+ - \Delta g_j^-\right>_{\Delta g_j \Delta N_j}$,
very large values of $\Delta g_j \Delta N_j$ can lead to relatively
large values of $\Delta g_j^+ - \Delta g_j^-$ of the opposite sign
reducing the aggregate return.

In addition to the distribution of the aggregate return, the
simulation does also agree with other properties of the empirical
data we found earlier in this paper. In Fig.~\ref{dg_rank.fig} and
\ref{dn_rank.fig} we also plotted the data from the simulation against
the rank according to the aggregate return $|G_j|$. For $\Delta N_j$
the simulation matches the empirical data very well, while in Figure
\ref{dg_rank.fig} we see that the simulated $\Delta g_j$ shows the same
dependence on the rank as the empirical data, but it is generally a
little larger than the real one except for the largest aggregate
returns, which might be due to the cutoff around 0.094 we used in the
simulation of the distribution of $\Delta g_j$.  We also find that the
role of $\Delta g_j^+ - \Delta g_j^-$ in determining large aggregate
returns is a little overestimated by our simulation, but the
simulation covers the main features of the empirical data although we
neglected many of the subtle relations between the different
quantities.

\section{Discussion and Conclusion}
	
Our results can be divided into two parts: First, we showed that the
movement of stock prices in intervals with a constant number of trades
can be understood as a diffusion process with a varying step
width. Here, Gaussian fluctuations of the number difference determine
the basic price movement, but the non-Gaussian large price changes are
due to changes in the tick return size coinciding with large number
differences at the same time. The large influence of the tick return
size is caused by its autocorrelations assuring that in a 100 tick
interval one can find many large tick returns so that the mean value
of the tick return can be large. In such intervals, the price change
in response to a trade is large, which is referred to as a period of
low liquidity. Thus, the diffusion process of stock returns depends
largely on fluctuations in the liquidity, in agreement with the
findings of previous works \cite{Plerou00, Farmer03, WeRo05, Weber06}.

In the second part of this paper, we found that the distribution of
aggregate returns can be reasonably approximated by simulating the
microscopic quantities mean tick return size and number difference
according to their empirically found distributions. A more accurate
agreement can be obtained by taking into account asymmetries in the
tick return size in positive and negative direction as well as
correlations between the different quantities.

To conclude, we found evidence that price fluctuations in intervals
with a constant number of trades can be described by a diffusion
process with a varying step width. The long-term autocorrelations in
the tick return size make sure that periods of low liquidity, where
the price change due to a trade is large, last long enough to cause
large aggregate returns in intervals with many trades. Our results
suggests that the power law distribution of aggregate returns might
not be universal but rather depends on a more complicated mechanism
which is a combination of the dynamics of the trading frequency, the
dynamics of the step width and the Gaussian process of the step
direction.

\section*{Acknowledgment}

I would like to thank Bernd Rosenow for collaboration in the initial
phase of the project, many helpful discussions, and a critical reading
of the manuscript. I also thank Fengzhong Wang for useful
conversations.

\end{document}